\documentclass[12pt]{iopart}

\usepackage[colorlinks, pdfborder={0 0 0}, plainpages=false,linkcolor=blue,citecolor=blue,urlcolor=magenta]{hyperref}
\usepackage{graphicx}
\usepackage{listings}
\usepackage{bm}
\usepackage{caption}
\usepackage{subcaption}
\usepackage[dvipsnames, usenames]{xcolor}
\usepackage[normalem]{ulem} 
\usepackage[numbers,square,comma,sort&compress]{natbib}

\begin{document}

\title[Numerical-relativity surrogate modeling with nearly extremal black-hole spins]{Numerical-relativity surrogate modeling with nearly extremal black-hole spins}


\author{Marissa Walker$^1$, Vijay Varma$^{2,\,*}$, Geoffrey Lovelace$^3$, Mark A. Scheel$^4$}
\address{$^1$ Christopher Newport University, Newport News, VA 23606}
\address{$^2$ Max Planck Institute for Gravitational Physics (Albert Einstein
Institute), D-14476 Potsdam, Germany}
\address{$^3$ Nicholas and Lee Begovich Center for Gravitational Wave Physics and Astronomy, California State University Fullerton, Fullerton, CA 92834, USA}
\address{$^4$ Theoretical Astrophysics, Walter Burke Institute for Theoretical Physics, California Institute of Technology, Pasadena, CA 91125, USA}
\address{*~Marie Curie Fellow}

\ead{marissa.walker@ligo.org}

\begin{abstract}
Numerical relativity (NR) simulations of binary black hole (BBH) systems provide
the most accurate gravitational wave predictions, but at a high computational
cost---especially when the black holes have nearly extremal spins (i.e. spins
near the theoretical upper limit) or very unequal masses. Recently, the
technique of Reduced Order Modeling (ROM) has enabled the construction of
``surrogate models'' trained on an existing set of NR waveforms. Surrogate
models enable the rapid computation of the gravitational waves emitted by BBHs.
Typically these models are used for interpolation to compute gravitational
waveforms for BBHs with mass ratios and spins within the bounds of the training
set. Because simulations with nearly extremal spins are so technically
challenging, surrogate models almost always rely on training sets with only
moderate spins. In this paper, we explore how well surrogate models can
extrapolate to nearly extremal spins when the training set only includes
moderate spins. For simplicity, we focus on one-dimensional surrogate models
trained on NR simulations of BBHs with equal masses and equal,
aligned spins. We assess the performance of the surrogate models at higher spin
magnitudes by calculating the mismatches between extrapolated surrogate model
waveforms and NR waveforms, by calculating the differences between extrapolated
and NR measurements of the remnant black-hole mass, and by testing how the
surrogate model improves as the training set extends to higher spins. We find
that while extrapolation in this one-dimensional case is viable for current
detector sensitivities, surrogate models for next-generation detectors should
use training sets that extend to nearly extremal spins.
\end{abstract}

\submitto{\CQG}
\maketitle

\section{Introduction}

Since LIGO's first detection in 2015~\cite{gw150914}, Advanced LIGO~\cite{aLIGO}
and Virgo~\cite{aVirgo} have observed dozens of gravitational waves from merging
black holes as the waves passed through Earth~\cite{gw150914, gwtc1,
    LIGOScientific:2020ibl, LIGOScientific:2021djp}. As the LIGO and Virgo detectors' sensitivities continue to improve, they will detect many more~\cite{observingscenarios}.

The observed gravitational-wave signals encode a wealth of information about
each source's nature and behavior. But extracting this information (``parameter
estimation'') is a computationally intensive process of comparing many model
waveforms with the data; Appendix E of Ref.~\cite{LIGOScientific:2021djp}, e.g.,
summarizes state-of-the-art parameter estimation techniques.

Binary black holes emit gravitational waves as the black holes spiral together,
merge into a new, larger remnant black hole that rings down into equilibrium.
Long before the time of merger (``inspiral'') and after the
time of merger (``ringdown''), approximate analytical approaches can give
accurate models of the gravitational waves; however, near the time of merger,
all analytic approximations fail; the waveforms can only be computed
using numerical relativity. (For a recent review of NR
techniques for modeling merging black holes, see, e.g.,
Ref.~\cite{Baumgarte:2021skc}.)

Developing high-accuracy waveform models will be especially important as
detectors improve their sensitivity, as signals will be recovered with higher
signal-to-noise ratios. Studies of the performance of current waveform models
with predicted third generation detector sensitivities show that there needs to
be an improvement of semi-analytic models' accuracy by three orders of magnitude, and
an improvement of NR accuracy by one order of magnitude
\cite{GWfuturedetectors_Purrer}. Numerical-relativity calculations are the most accurate way to compute the gravitational waves emitted by
merging black holes, but they are too computationally expensive to feasibly
yield the large number of waveforms that gravitational-wave inference methods
require. One approach to overcoming this challenge is the construction of
approximate, semi-analytic  models~\cite{Ossokine:2020kjp, Pratten:2020ceb,
Estelles:2021gvs, Akcay:2020qrj, Cotesta:2018fcv, Garcia-Quiros:2020qpx,
Estelles:2020twz, Nagar:2020pcj} that are calibrated to and validated against
catalogs of numerical-relativity waveforms. For instance, two such
models, SEOBNRv4PHM~\cite{Ossokine:2020kjp} and
IMRPhenomXPHM~\cite{Pratten:2020ceb}, were used to infer the properties of binary black holes in the most recent Gravitational-Wave Transient Catalog~\cite{LIGOScientific:2021djp}.

An alternative approach is to create a stand-in, or ``surrogate,''
model~\cite{Field:2013cfa} that effectively interpolates an existing set of NR waveforms (the ``training set''), each with different binary parameters, to produce a waveform with a desired set of binary parameters. After a surrogate model
is created, it can quickly evaluate a gravitational
waveform whose parameters are contained within the most extreme parameters of
the training set. Surrogate models can also extrapolate beyond their training
set, but the results become progressively less accurate. Surrogate models
trained directly on NR simulations~\cite{Varma2019_NRSur7dq4, NRHybSur3dq8,
Blackman:2017pcm} have been used alongside approximate, semi-analytic models in
recent analyses of LIGO-Virgo gravitational wave
observations~\cite{LIGOScientific:2020ibl, LIGOScientific:2020iuh,
LIGOScientific:2020stg, Varma:2022pld, Varma:2021xbh, Varma:2021csh,
Varma:2020nbm, Kumar:2018hml, Islam:2020reh}. Surrogate models have an advantage in
accuracy over approximate, semi-analytic models when comparing with numerical
relativity, but they also have some disadvantages: they are only guaranteed
to be accurate within the parameter space spanned by the training set, and
their length (i.e., number of orbits) is limited to the length of the training waveforms, meaning they
can only span a detector's sensitive frequency band for binaries with
sufficiently high masses (although, this limitation has been lifted for
aligned-spin BBHs~\cite{NRHybSur3dq8}).

One particularly interesting region of BBH parameter space is binaries
containing a black hole spinning nearly as rapidly as theoretically possible;
observational evidence (e.g.~\cite{gou2011extreme}) suggests that such black
holes exist and thus might be among the merging black holes that
gravitational-wave detectors will observe. Black hole spins are described in this
paper using the dimensionless spin $\chi = S/M^2$ (where $S$ is the
spin angular momentum and $M$ is the Christodoulou mass of the black hole), which ranges from -1 to 1, with the most rapidly spinning black holes having spins close to $\pm$1.
Current surrogate models based on NR~\cite{Varma2019_NRSur7dq4, NRHybSur3dq8,
Blackman:2017pcm} are typically trained on waveforms
containing black holes with spins up to $\chi=0.8$. This limitation follows from
the practical difficulty in modeling merging black holes with nearly
extremal spins~\cite{Scheel:2014ina, Healy:2017vuz}. 

Ideally, waveform models would cover
the entire parameter space, but creating a training set of high spin and high
mass ratio waveforms is computationally challenging and expensive. Using current
surrogate models to predict BBH waveforms including a rapidly spinning (greater
than 0.8) black hole requires extrapolation beyond the training parameters in
spin. This work explores two questions within the context of a simplified
one-dimensional surrogate model: the accuracy of such extrapolation compared
with NR, and the potential for improving high spin models by expanding the
training set.

The rest of this paper is organized as follows: Section~\ref{sec:methods}
describes the methods used for constructing surrogate models, summarizes the
process of producing the numerical relativity waveforms that were used, and
estimates the level of accuracy needed for waveforms.
Section~\ref{sec:waveforms} presents the one-dimensional waveform surrogate
models and analysis of results for extrapolation to high spin.
Section~\ref{sec:remnant} presents the results of using the same surrogate
modeling process to predict the mass and spin of the remnant black hole.
Section~\ref{sec:conclusion} summarizes the findings and future outlook of the
work. Note that throughout this paper we use units $G = c = 1$.

\section{Methods}\label{sec:methods}

\subsection{Surrogate Modeling}
Surrogate models trained on numerical relativity waveforms are a fast way of
constructing the many waveforms needed for parameter estimation. Several
numerical relativity surrogate waveform models exist covering various ranges of
parameters. Generically, BBHs can be described using seven dimensions: three dimensions to describe the spin of each black hole, and one to describe the mass ratio between the two black holes ($q = M_1/M_2$, where we choose the convention that $M_1$ is the heavier object, so $q\geq 1$). Two existing surrogate models will be used to show the capabilities
of current models for high spin black holes. One model is the NRSur7dq2
surrogate model, which covers generically spinning black holes (including
precession following from spins misaligned with the orbital angular momentum)
with mass ratios up to 2 and spin magnitudes $\chi_{1,2}\leq0.8$ \cite{Blackman:2017pcm}. A recent update to this model,
NRSur7dq4 \cite{Varma2019_NRSur7dq4}, expands the mass ratio range to 4, but
results in this paper use NRSur7dq2 since here we treat only the equal-mass case. Another model NRHybSur3dq8 \cite{NRHybSur3dq8}, includes mass ratios up to 8, but only for
non-precessing spins. This three dimensional model is also hybridized with
analytic models to create waveforms longer than the numerical relativity
simulations by combining the early inspiral from analytic models with the final
portion near and after the time of merger from numerical relativity.

The surrogates produced in this paper use a similar method to those previous
models, as described in
\cite{NRHybSur3dq8}. The initial step in the process is gathering the
numerical relativity waveforms of the training set from the SXS Catalog
(Sec.~\ref{sec:NR}). The waveforms were chosen to span the desired parameter space, and to take advantage of waveforms that already exist in the catalog. Specifically, we used waveforms where the two black holes had equal mass (mass ratio $q = 1.0$) and equal, aligned or anti-aligned spin. Different ranges of spin were used for different surrogates, as described in Sec.~\ref{sec:surrogates}.

The waveforms are time-shifted such that the peak of the
signal occurs at time t=0, and the beginning of the waves are truncated such
that all the waveforms begin at the same time and do not include the initial burst of
spurious gravitational waves at the beginning of each simulation (these spurious waves are caused by
limitations in the methods used to construct initial data). The surrogates
presented in this paper use an initial time of -1000M, at which time the frame is aligned so that the initial orbital phase is zero. For
the surrogates presented in this paper, since all of the binary black holes have mass ratio $q=1$ and have spin that is aligned or anti-aligned with the orbital axis, the $l=m= 2$
spin-weighted spherical harmonic modes of gravitational radiation will
dominate~\cite{Varma:2016dnf, Varma:2014jxa, Capano:2013raa,
Shaik:2019dym}. For this reason, we have only included the $l = m= 2$ mode. Surrogate
models work better when the training data consists of slowly varying functions, so the
training waveforms are decomposed into separate data pieces that are more
slowly varying in time. In this case, the data pieces used are the amplitude
and phase of the (2,2) mode of the waveform.  For each data piece a
separate surrogate is constructed, and then the final waveform is modeled by
combining the different data pieces.

For each surrogate, reduced basis decomposition is used to create a linear basis to represent the training set \cite{PhysRevLett.106.221102}. Additional basis functions are added until all the waveforms can be represented with errors below a given tolerance \cite{Blackman:2017dfb}. We used an amplitude basis tolerance $10^{-3}$ and a phase basis tolerance of $10^{-2}$. A subset of the times is then chosen to be time nodes, using a greedy process to determine the most representative times for the waveform data pieces so that the data can be interpolated between the nodes to construct the full waveform \cite{BARRAULT2004667, 1534-0392_2009_1_383, refId0}. Finally, for each time node, a fit is created for the
data piece across the different parameters of the binary black hole. In the
case of the surrogates created here, the only varying parameter is the spin of the black holes. Since each black hole pair has equal spins ($\chi_{1}=\chi_{2}$), the surrogates are one-dimensional, and the fit parameter is the spin. 
The fits are constructed using the forward-stepwise greedy parametric fitting method described in App. A of Ref.~\cite{Blackman:2017dfb} allowing up to fourth order monomials in the basis functions. To avoid overfitting, a cross-validation step is taken where 10 trial fits are performed, leaving out one waveform in each one to validate the fit. 

We also created a surrogate to interpolate the final mass and spin of the remnant black
hole as a function of the initial spin. We followed the approach of~\cite{Varma:2018aht, Varma2019_NRSur7dq4} but ignored the recoil velocity because it is zero due to the symmetries of equal mass, equal spin BBHs.

\subsection{Numerical relativity waveforms}\label{sec:NR}
We train our surrogate models using previously published numerical-relativity
waveforms from the Simulating eXtreme Spacetimes (SXS)
catalog~\cite{Mroue:2013xna, SXScatalog2019}. Here, we briefly summarize the
techniques that SpEC uses to simulate binary black holes; for a recent, more
detailed discussion of these techniques, see Sec.~II of
Ref.~\cite{SXScatalog2019} and the references therein.

Each waveform was computed using the Spectral Einstein Code
(SpEC)~\cite{SpECwebsite}. SpEC constructs binary-black-hole initial data in
quasi-equilibrium by solving the eXtended Conformal Thin Sandwich (XCTS)
formulation of the Einstein constraint equations~\cite{York:1998hy,
Pfeiffer:2002iy} using excision boundary conditions~\cite{Cook:2004kt} and free
data based on a weighted superposition of two Kerr-Schild black
holes~\cite{Lovelace:2008tw}. SpEC then evolves the initial data by solving the
generalized-harmonic formulation of the Einstein evolution
equations~\cite{Lindblom:2005qh} using a pseudospectral
approach~\cite{Szilagyi:2009qz} with a constraint-preserving boundary condition
applied on the outer boundary~\cite{Rinne:2006vv}.

The singularities are excised from the computational domain, with a control
system dynamically transforming the computational grid to conform to the black
holes' apparent horizons as they move and deform while ensuring that the
excision boundaries require no boundary condition (by ensuring that the
characteristic speeds are all outgoing)~\cite{Hemberger:2012jz,
Ossokine:2013zga}. When the black-hole spins are nearly extremal, the excision
problem becomes especially delicate: as the spin approaches the theoretical
maximum, the excision surface must conform more and more precisely to the
apparent horizon---while remaining inside of it---to avoid incoming
characteristic speeds. Recent improvements in SpEC's
techniques~\cite{Scheel:2014ina} have enabled SpEC to simulated binary black
holes with spins near the theoretical maximum. Specifically, SpEC has simulated
merging black holes with dimensionless spin $\chi$ as high as 0.998~\cite{SXScatalog2019}. 

\subsection{Surrogate accuracy}
The surrogate models must be sufficiently accurate---but how accurate is
accurate enough? Ensuring that current and future gravitational wave detectors
can precisely determine the physical properties of the loudest black hole
mergers that they observe and can precisely compare the waves with general
relativity's predictions \cite{testingGRgwtc1} require model waveforms whose
numerical errors do not exceed the experimental uncertainty. As an observed
gravitational wave's signal-to-noise ratio increases, the theoretical models
must become correspondingly more accurate.

One way to measure the error in a waveform is the \textit{mismatch} between the
waveform and a fiducial waveform. The mismatch between two complex waveforms $h_1$ and $h_2$ is calculated in the time domain, using:
\begin{equation}
    \mathcal{M} = 1 - \frac{\langle h_1,h_2 \rangle}{\sqrt{\langle h_1,h_1 \rangle \langle h_2, h_2 \rangle}},
\end{equation}
where the inner product between waveforms is defined by 
\begin{equation}
    \langle h_1, h_2 \rangle = \Big| \int_{t_{min}}^{t_{max}} h_1(t) h_2^*(t)dt\Big|.
\end{equation}
The mismatch is analogous to taking one minus the dot product between two unit vectors. If the vectors (waveforms) are identical, the mismatch will be zero.

As a crude estimate, two template waveforms may be considered to be indistinguishable if mismatch between them is low enough that the following
condition is met~\cite{lindblom}:
\begin{equation}
\mathcal{M}< \frac{\mathcal{D}}{2\rho ^2}
\end{equation}
where $\mathcal{M}$ is the mismatch between the two waveforms, $\mathcal{D}$ is
the number of parameters used to describe the detection (which is 8 for
generically spinning black holes assuming circular orbits), and $\rho$ is the signal-to-noise ratio
(SNR) of the detection. Therefore as SNRs increase, the acceptable level of
error in the theoretical waveforms decreases. As an example, in the first three observing runs of the current detectors, the loudest gravitational wave signal
from a binary black hole system was GW$200129\_065458$, with SNR of 26.5 combined between three detectors~\cite{LIGOScientific:2021djp}. Using
the mismatch condition above, a mismatch of $\mathcal{M}< 5.7 \times 10^{-3}$ would be sufficient. However, if an optimally oriented binary black hole signal with the
same properties and distance as the first BBH detection, GW150914, were to happen when LIGO reaches
its design sensitivity, the SNR
could be nearly 100 for a single detector \cite{gw150914detectors}, corresponding to a required mismatch of $4 \times 10^{-4}$.

Binary black hole observations could be detected with SNR up to several hundred
or even 1000 with third-generation detectors \cite{ET_Hild_2011,
3g_sensitivity_Hall_2019}. With an SNR of 1000, the mismatch condition goes
down to $\mathcal{M}< 4 \times 10^{-6}$. It is therefore critical to push the
boundaries of accuracy in waveform modeling, as well as the models' efficiency
to produce large numbers of waveforms for parameter estimation.

\section{Surrogate modeling for high spin}\label{sec:surrogates}

We used existing NR waveforms from the SXS catalog \cite{SXScatalog2019} to
create and test one-dimensional surrogate models, first to examine the effects
of extrapolation beyond current surrogate spin levels and then to test the
extrapolation with different training sets. All surrogate models used in these
tests follow a similar method to that described in \cite{NRHybSur3dq8}.

\subsection{Case numbers and information about waveforms}
All of the waveforms used in the training and validation sets for the
one-dimensional surrogate models presented in this paper are publicly available
in the SXS Waveform Catalog. Table~\ref{wavetable} summarizes the configurations
used. All of these waveforms are binary black hole systems with equal masses,
and equal spins either aligned or anti-aligned with the orbital angular
momentum. Each binary has initial parameters tuned~\cite{Buonanno:2010yk} so that the initial orbital eccentricity at most of order $10^{-3}$.
\begin{center}
\begin{table}
\begin{tabular}{||c c c c c||}
 \hline
Name & $\chi_{1z}$ & $\chi_{2z}$ & $e$ & $N_{\rm orbits}$\\ [0.5ex]
 \hline\hline
SXS:BBH:0180 & $-3.67 \times 10^{-9}$ & $-2.157 \times 10^{-9}$ & $5.110 \times 10^{-5}$ & 28.18 \\
SXS:BBH:0149 &  -0.2000 & -0.2000 & $1.604 \times 10^{-4}$ & 17.12 \\
SXS:BBH:0150 &  0.2000 & 0.2000 & $2.714 \times 10^{-4}$ & 19.82 \\
SXS:BBH:0148 &  -0.4376 & -0.4376 & $<3.500 \times 10^{-5}$ & 15.45 \\
SXS:BBH:1122 &  0.4376 &  0.4376  & $3.727 \times 10^{-4}$ & 21.53 \\
SXS:BBH:0151 &   -0.5999 &  -0.5999 & $<4.800 \times 10^{-4}$ & 14.48 \\
SXS:BBH:0152 &  0.6000 &  0.6000 & $4.272 \times 10^{-4}$ & 22.64 \\
SXS:BBH:0154 &  -0.7998  & -0.7998 & $<6.400 \times 10^{-4}$ & 13.24 \\
SXS:BBH:0155 &  0.7999 & 0.7999 & $5.051 \times 10^{-4}$  & 24.09 \\
SXS:BBH:0153 &  0.8498 & 0.8498 & $8.694 \times 10^{-4}$ & 24.49 \\
SXS:BBH:0159 &  -0.8996 & -0.8996 & $<8.100 \times 10^{-4}$ & 12.67 \\
SXS:BBH:0160 &  0.8997 & 0.8997 & $4.442 \times 10^{-4}$ & 24.83 \\
SXS:BBH:0156 &  -0.9490 & -0.9490 & $7.671 \times 10^{-4}$ & 12.42 \\
SXS:BBH:0157 &  0.9496 & 0.9496 & $1.483 \times 10^{-4}$ & 25.15 \\
SXS:BBH:1137 &  -0.9692 & -0.9692 & $4.313 \times 10^{-4}$ & 12.19 \\
SXS:BBH:0172 &  0.9794 & 0.9794 & $1.128 \times 10^{-3}$ & 25.35 \\
SXS:BBH:0177 &  0.9893 & 0.9893 & $<2.000 \times 10^{-3}$& 25.40 \\[1ex]
  \hline

 \hline
\end{tabular}
\caption{\label{wavetable} The numerical-relativity waveforms used in the
    surrogate models and testing shown in this paper. Each simulation models
    merging binary black holes with equal masses ($q=1.000$) and equal spins aligned or
    anti-aligned with the orbital angular momentum. Shown are the simulation
    name, the spin of each black hole, an estimate of the orbital eccentricity, and the number of orbits $N_{\rm orbits}$ simulated before merger.}
\end{table}
\end{center}

\subsection{Extrapolation beyond training set}\label{sec:waveforms}
\begin{figure*}[ht]
\begin{center}
\includegraphics[width=0.8\linewidth]{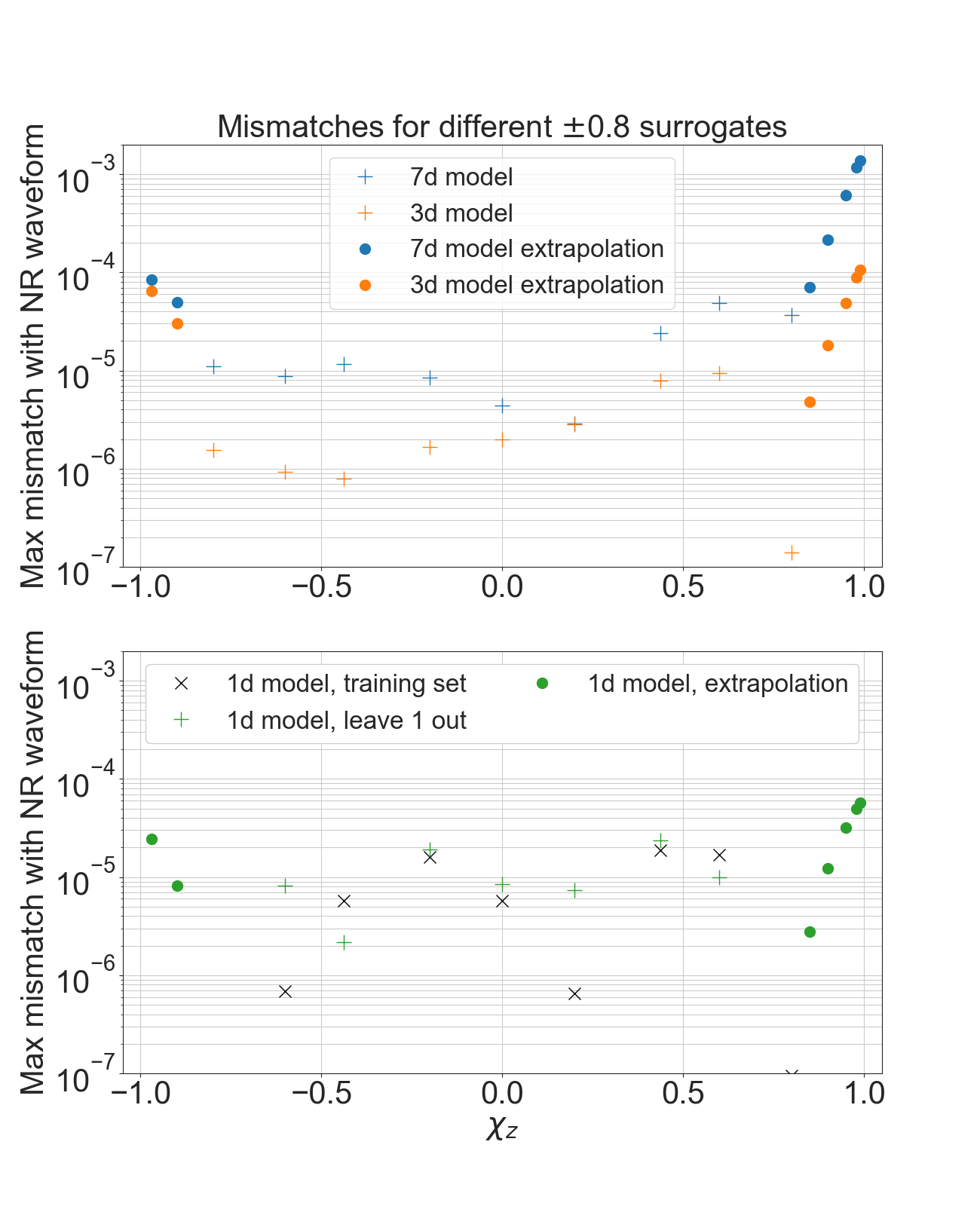}
\caption{Above: Mismatches between numerical relativity and surrogate evaluation for various spins for a set of equal mass, equal and aligned spin black holes, using two existing surrogate models, NRSur7dq2 and NRHybSur3dq8, using test cases from the NR waveforms listed in Table~\ref{wavetable}.  Below: Mismatches between the NR simulation and surrogate evaluation using a one-dimensional surrogate model in spin. The surrogate model was trained on nine simulations of equal mass and
equal aligned spin, with spins from -0.8 to 0.8. A leave-one-out
validation study was also performed, with surrogate models trained on the same set except for one of the waveforms. For the validation cases, the corner case waveforms (spin -0.8 and 0.8) were left in the model each time.
In all cases in both plots, models were trained spin magnitudes up to 0.8. As the models are used to extrapolate farther from that training set range, the mismatches get worse.
}
\label{fig:mismatches_diffsurs}
\end{center}
\end{figure*}

First to determine the applicability of current surrogate models to the high
spin parameter space, existing models can be extrapolated beyond their optimal
(training set) range to create waveforms for rapidly spinning binary black
holes. Several different parameters were chosen for this prediction based on
existing numerical simulations within the SXS catalog, so that mismatch
comparisons could show how accurate this extrapolation is. Specifically, we
chose test cases that have equal aligned spin and equal mass, so the only
parameter that varies is the magnitude of the spin. The top plot in Figure
\ref{fig:mismatches_diffsurs} demonstrates the capability of two existing
models to accurately model systems within the parameter space of their training
set, as well as the effects of extrapolating beyond the training set range.

As might be expected, as the models are extrapolated further beyond the
training set range, the mismatches with NR simulations get worse. However, even
the highest spin case ($\chi = 0.9893$) is still predicted with a mismatch of
less than the mismatch criteria for current detectors. While this result is
promising for the applicability of current surrogate models across the entire
spin parameter space (at least in the equal mass and equal, aligned spin case),
improvements in the models are needed to meet the mismatch criteria for future
detectors.

Next, to create a simplified model solely for the purposes of exploring the effects of spin, a one-dimensional surrogate model was created from a set of waveforms with equal mass and equal, aligned spin. Specifically, the waveforms used were the first nine in Table \ref{wavetable} with spin $\chi \leq \pm 0.8$. The lower plot in Figure \ref{fig:mismatches_diffsurs} shows the mismatches between the NR simulations and the  evaluations of this simplified surrogate model. Since the test cases with $\chi \leq \pm 0.8$ were also part of the training set used to create the model, the mismatches for those cases are also shown with a ``leave one out test", where the model was recreated by leaving out only that test case from the training set and then doing the mismatch test on that case. These tests showed that this simplified model is roughly comparable or a bit better than existing models, so we can use this technique to explore different types of training sets.

\subsection{Varying the spin in the training sets}

We then created several different one-dimensional surrogate models, each with a
different number of training set waveforms that include a different range of
spins. Each training set used a subset of the waveforms listed in Table
\ref{wavetable}. These sets were created by choosing all of the cases within the range of spins from $\pm |\chi_{max}|$, where the maximum spin $\chi_{max}$ was different for each training set. For example, the training set for the $|\chi_{max}| = 0.2$ model simply consisted of the first three cases in the table, with spins of 0 and $\pm0.2$. The next model created (with $|\chi_{max}| = 0.4376$) was trained on the first five cases, including the original three plus the two higher spin cases. Each set therefore contained two additional training cases from the previous set.

\begin{figure*}[ht]
\begin{center}
\includegraphics[width=0.9\linewidth]{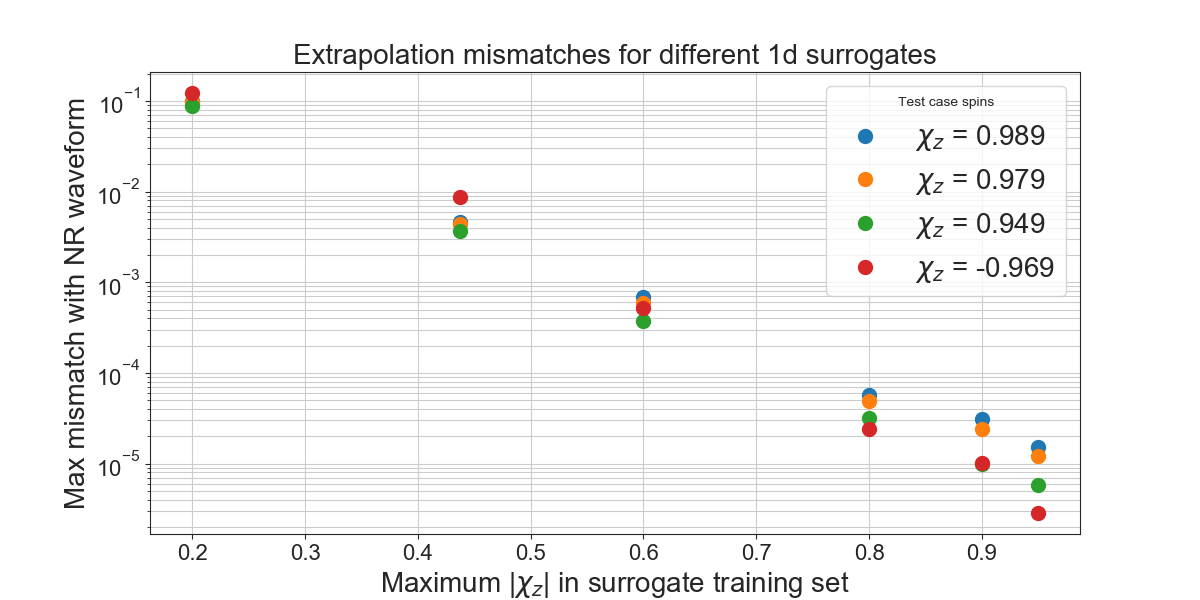}
\caption{Mismatches with NR simulations and surrogate evaluation of four high spin test cases, for surrogates with different training sets. Each training set encompasses a range of spins from $\pm|\chi_z|$. As the maximum spin included in the training set increases, the extrapolation mismatch gets smaller. (Note that for the 0.95 spin case, the waveform was a validation case as well as a training case for the surrogate with maximum spin of 0.95.)}
\label{fig:extrapmismatches_diffsurs}
\end{center}
\end{figure*}

For each different training set, we tested how well the model
extrapolated to the most rapidly spinning black holes. Specifically, we used
spins of magnitude 0.949 and greater to test the extrapolation of the models.
Figure \ref{fig:extrapmismatches_diffsurs} shows the mismatch test results from
these different surrogates. As expected, when only using three low spin
waveforms up to maximum spin magnitude of 0.2, the model has huge errors when
extrapolating to high spin. With each additional set of corner points added,
however, the mismatches decrease. Even only including up to 0.6 spins in the
training sets reduces the extrapolation mismatches for all of the validation
cases to less than 1e-3. Two of the surrogates tested and shown in Figure
\ref{fig:extrapmismatches_diffsurs} have an expanded training set beyond
$|\chi|=\pm 0.8$: one includes up to $|\chi|=\pm 0.9$, and the other up to
$|\chi|=\pm 0.95$. We see that the trend continues, with these surrogates
trained on higher spin performing better in their predictions of the high spin
waveforms.

\begin{figure*}[ht]
\begin{center}
\includegraphics[width=0.9\linewidth]{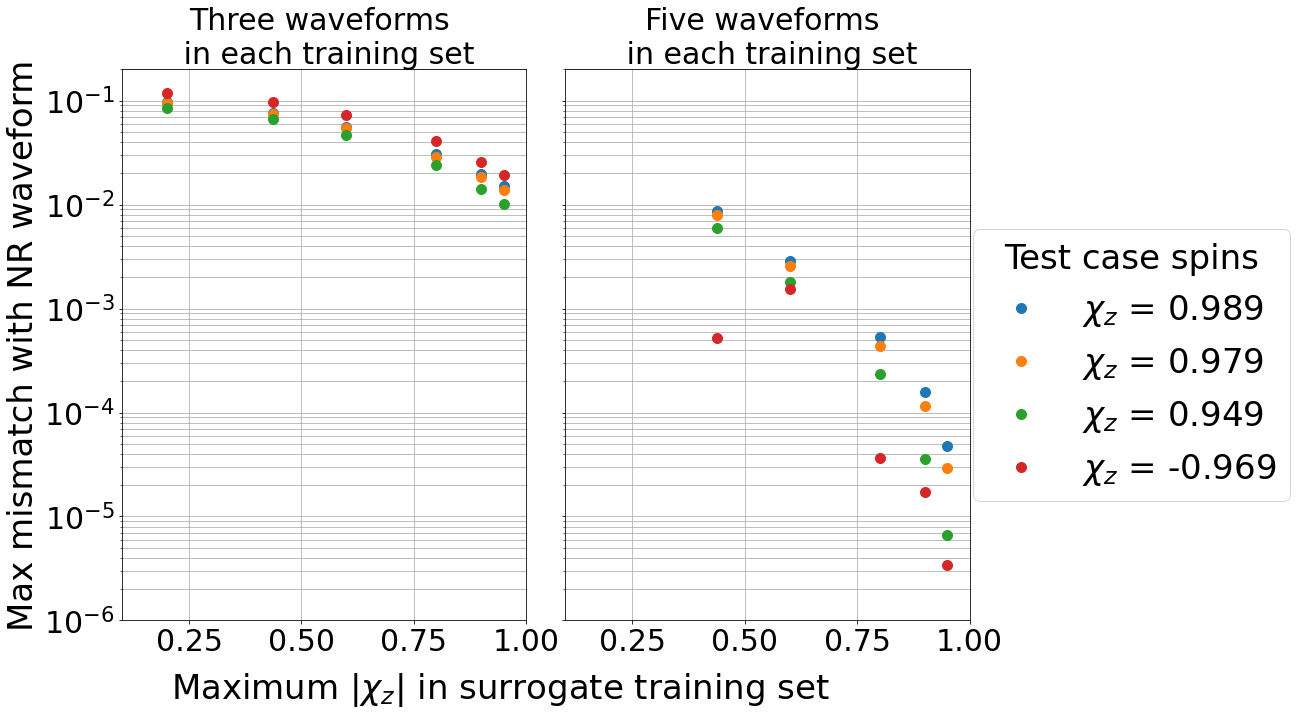}
\caption{Both plots above show the results of extrapolating to high spin waveforms using surrogates trained on different sets, similar to Figure \ref{fig:extrapmismatches_diffsurs}. However, in these cases, the number of waveforms is fixed, and only the maximum spins change. In the plot on the left, each training set consists of only three waveforms: 0 and $\pm |\chi_{max}|$. With only three points in the training set, the mismatches are all fairly high, even when $|\chi_{max}|$ is 0.95. However, there is an improvement as the spin increases. On the right hand side, each training set consists of five waveforms: 0, $\pm 0.2$, and $\pm |\chi_{max}|$. While the mismatches are still not quite as good as when the training set size increased at every step, there is a significant improvement with increasing the maximum spin in the training set.} 
\label{fig:extrapmismatches_diffsurs_3and5}
\end{center}
\end{figure*}

In the tests shown in Figure~\ref{fig:extrapmismatches_diffsurs}, the number of waveforms in the training set changed with each different surrogate, as well as the maximum spin. In order to isolate the effect of the spin in the training set, we repeated the procedure of creating multiple surrogates with different maximum spins but the same number of waveforms in each training set. First, we created multiple surrogates with only three waveforms in each training set: 0 and $|\pm \chi_{max}|$, where the maximum spin changed each time. When these surrogates are extrapolated to evaluate the highest spin waveforms, there is still a clear effect of lower mismatches when the maximum spin is increased. However, the mismatches are all significantly higher, with the lowest mismatches being about $10^{-2}$. Next, we used five waveforms in each training set: 0, $\pm 0.2$, and $\pm |\chi_{max}|$. In these cases, the mismatches are significantly lower, though not quite as low as the surrogates where all the possible waveforms between $\pm |\chi_{max}|$ were used. The mismatch results from these tests are shown in Figure~\ref{fig:extrapmismatches_diffsurs_3and5}.

\subsection{Remnant Mass Surrogate Models}\label{sec:remnant}
A similar process can be followed to develop a model not for the full predicted
gravitational waveform, but simply for the properties of the remnant black
hole, specifically the mass and spin of the remnant.

\begin{figure}
     \centering
\includegraphics[width=0.7\textwidth]{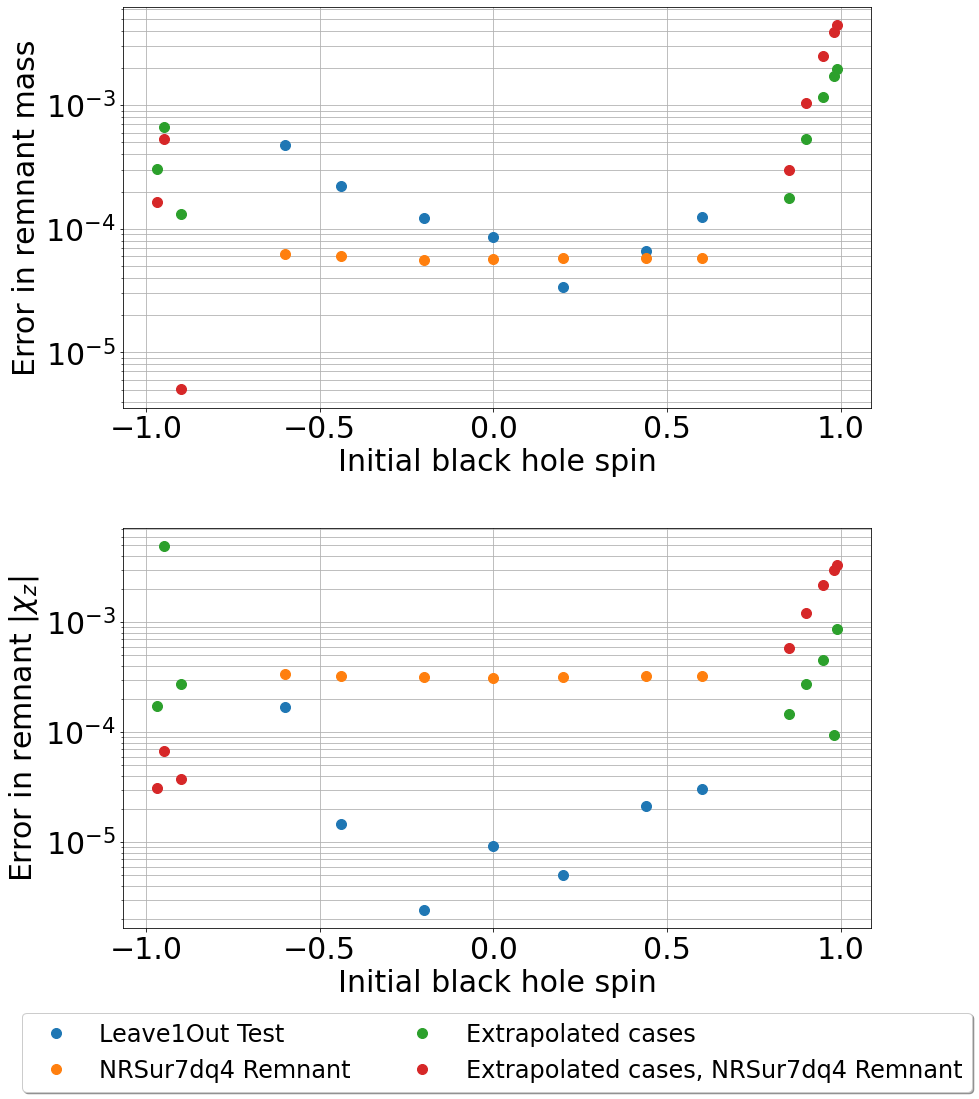}
\caption{
 Differences between the remnant masses and spins computed from the NR simulations in Table \ref{wavetable}, and the remnant masses and spins computed by two remnant surrogate models. Blue: Leave-one-out tests with the 1D surrogate,
where that particular waveform was left out of the training set and used only
for testing. Orange: Cases from the table with spins $|\chi|\leq 0.8$ compared to the surrogate model NRSur7dq4Remnant. Green: Extrapolated predictions with the 1d model. Red:
Extrapolated predictions from the NRSur7dq4Remnant model.
}
\label{fig:remnantmassandspin}
\end{figure}

Just as described above, we first trained a remnant surrogate on the nine BBHs
in Table \ref{wavetable} with spin magnitudes less than or equal to 0.8, using the same surrogate model fitting procedure as for the waveform model. To
test the capabilities of this surrogate within the realm of the training set,
we performed leave-one-out tests as described above. Then we tested the
surrogate model's ability to extrapolate to high spin. As a reference, the
NRSur7dq4Remnant surrogate model (described in \cite{Varma2019_NRSur7dq4}) was
also evaluated at the same points. The surrogate model errors, or the differences in the surrogate predictions from the NR calculations, both for the interpolated and extrapolated cases, are shown in Figure \ref{fig:remnantmassandspin}. Similar to the waveform surrogate models, most of the
extrapolated cases have a higher error than those within the training set
range, with errors increasing as the extrapolation gets further away from the
training range.

\begin{figure}
     \centering
\includegraphics[width=0.7\textwidth]{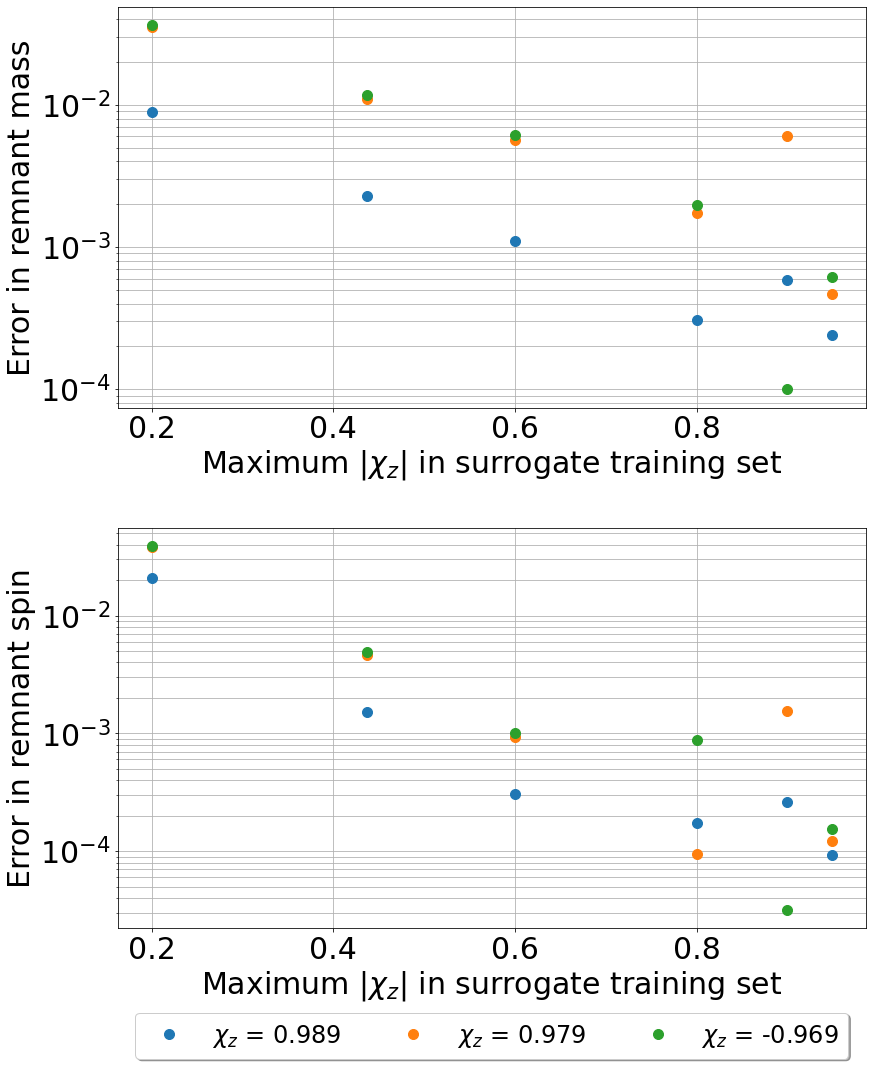}
\caption{Surrogate model errors (differences from NR calculations) in the extrapolated remnant mass and spin predicted by
    different one-dimensional surrogate models trained with different training sets with different maximum spins.
    Similar to the waveform surrogate, there is a clear overall pattern of
    decreasing error as the training set is expanded to include higher spin
simulations.
}
        \label{fig:remnantmassandspin_extrap}
\end{figure}

Similar to the waveform surrogates, we prepared several different one-dimensional surrogates
trained on different ranges of spin and tested how well each of these performed
on extrapolating to extreme spin cases. The differences between these predictions and the NR calculations are
shown in Figure \ref{fig:remnantmassandspin_extrap}. In these surrogates, all of the waveforms from Table~\ref{wavetable} between $\pm |\chi_{max}|$ were used for each training set. Once again, the trend matches what that of the waveform surrogate studies: expanding the training set to include the higher spin BBHs improves the extrapolation to high spin.

\section{Conclusions}\label{sec:conclusion}
One main conclusion from this study is that extrapolation from surrogate models
might actually be viable to some extent for current detector sensitivities.
There are of course limitations in the study, since we were only extrapolating
in one dimension. Additionally, higher order modes beyond the $l=m=2$ mode have
not been considered. Further study would need to be conducted to test
extrapolation in a more generic case; for example a BBH with unequal masses.
However, it is promising that for the simplest cases shown here, mismatches for
extrapolating to rapid spins may already be suitable for the SNRs expected to
be detected from current detectors, and possibly even for future detectors.
This finding is consistent with tests of the surrogate model NRSur7dq4, which
was trained on mass ratios up to 4 but was shown to produce accurate
predictions of BBHs up to mass ratios of 6 \cite{Varma2019_NRSur7dq4}.

To reach mismatches of the order of $10^{-6}$ (which is sufficient for
indistinguishability for SNR 1000 observations) across the parameter space,
however, the models would need to be improved. This study shows that surrogate extrapolation in spin will continue to improve if we include higher spin waveforms in training sets. When we have higher signal-to-noise observations we
will want more accurate models and expanded parameter space. Therefore it is
worthwhile to continue to conduct high spin simulations, especially with
unequal masses, in order to expand the applicability of surrogate models.
Advances in numerical relativity simulations as well as surrogate modeling
methods remain important for progressing towards this increased level of
accuracy.

\ack{
This work was supported in part by NSF award PHY-2011975 at Christopher Newport
University; by NSF awards PHY-1654359 and PHY-1606522 and by the Dan Black
Family Trust and Nicholas and Lee Begovich at Cal State Fullerton.
V.V. acknowledges funding from the European Union's Horizon 2020 research and
innovation program under the Marie Skłodowska-Curie grant agreement No.~896869.
M.S. acknowledges funding from the Sherman Fairchild Foundation
and by NSF Grants No. PHY-1708212, No. PHY-1708213, and No. OAC-1931266
at Caltech.
}


\bibliographystyle{iopart_num}
\bibliography{references}

\end{document}